\documentclass[preprints,article,accept,moreauthors,pdftex]{Definitions/mdpi} 

\usepackage{bm}
\usepackage{listings}
\usepackage{subfigure}
\usepackage{amsmath}
\usepackage{amssymb}
\usepackage{booktabs}
\firstpage{1} 
\pubvolume{xx}
\issuenum{1}
\articlenumber{1}
\pubyear{2020}
\copyrightyear{2020}
\history{Received: date; Accepted: date; Published: date}

\newcommand{\argmin}{\mathop{\mathrm{argmin}}}

\newcommand{\coshm}{\mathop{\mathrm{coshm}}}
\newcommand{\sinhm}{\mathop{\mathrm{sinhm}}}
\Title{An arbitrage-free interpolation of class $\mathcal{C}^2$ for option prices}
\Author{Fabien {Le Floc'h}}
\AuthorNames{Fabien Le Floc'h}
\address{Delft Institute of Applied Mathematics, TU Delft, Delft, The Netherlands}
\corres{Correspondence: F.L.Y.LeFloch@tudelft.nl}
\abstract{This paper presents simple formulae for the local variance gamma model of Carr and Nadtochiy, extended with a piecewise-linear local variance function. The new formulae allow to calibrate the model efficiently to market option quotes. On a small set of quotes, exact calibration is achieved under one millisecond. This effectively results in an arbitrage-free interpolation of class $\mathcal{C}^2$. The paper proposes a good regularization when the quotes are noisy. Finally, it puts in evidence an issue of the model at-the-money, which is also present in the related one-step finite difference technique of Andreasen and Huge, and gives two solutions for it.  }

\begin{document}
	
\section{Introduction}
The financial markets provide option prices for a discrete set of strike prices and maturity dates. In order to price over-the-counter vanilla options with different strikes, or to hedge complex derivatives with vanilla options, it is useful to have a continuous arbitrage-free representation of the option prices, or equivalently of their implied volatilities.  For example, the variance swap replication of Carr and Madan consists in integrating a specific function over a continuum of vanilla put and call option prices \citep{carr2001towards,carr2008robust}. More generally, \citet{breeden1978prices} have shown that any path-independent claim can be valued by integrating over the probability density implied by market option prices. An arbitrage-free representation is also particularly important for the Dupire local volatility model \citep{dupire1994pricing}, where arbitrage will translate to a negative local variance. A option price representation of class $\mathcal{C}^2$ is also key to guarantee the second-order convergence of numerical schemes applied to the Dupire partial differential equation, commonly used to price exotic financial derivative contracts. In this paper, we describe a new technique to interpolate the market option prices in an arbitrage-free manner, with very high accuracy.

A rudimentary, but popular representation is to interpolate market implied volatilities with a cubic spline across option strikes. Unfortunately this may not be arbitrage-free as it does not preserve the convexity of option prices in general. A typical convex interpolation of the call option prices by quadratic splines or rational splines is also not satisfactory in general since it may generate unrealistic oscillations in the corresponding implied volatilities, as evidenced in \citep{jackel2014clamping}.  \citet{kahale2004arbitrage} designs an arbitrage-free interpolation of the call option prices, which however requires convex input quotes, employs two embedded non-linear minimizations, and it is not proven that the algorithm for the interpolation function of class $\mathcal{C}^2$ converges. 


More recently, \citet{andreasen2011volatility} have proposed to calibrate the discrete piecewise constant local volatility corresponding to a single-step finite difference discretization of the forward Dupire equation. In their representation of the local volatility, the authors use as many constants as the number of market option strikes for an optimal fit. It is thus sometimes considered to be "non-parametric". Their technique works well in general but requires some care around the choice of discretization grid: it must be sufficiently dense so that two market strikes do not fall in between the same consecutive grid nodes, and sufficiently wide to properly model the boundary behaviour. Those two requirements complicate, and slow down the non-linear optimization involved in the technique. Furthermore the output is a discrete set of option prices, which, while relatively dense, must still be interpolated carefully to obtain the price of options whose strike falls in between grid nodes.

\citet{lefloch2019model} derived a specific B-spline collocation to fit the market option prices, while ensuring the arbitrage-free property at the same time. While the fit is quite good in general, it may not be applicable to interpolate the original quotes with high accuracy. For example, input quotes may already be smoothed out if they stem from a prior model, or from a market data broker, or from another system in the bank. In those cases, it is desirable to use a nearly exact interpolation.

Here, we extend the local variance gamma arbitrage-free interpolation of class $\mathcal{C}^1$ presented in \citep{carr2017local}. Instead of a piecewise-constant parametrization for the variance function, we use a piecewise-linear representation. The resulting partial differential
difference equations can still be solved explicitly and the outcome is an interpolation of class $\mathcal{C}^2$. Being based on the local variance gamma model, we find that it behaves properly, even on challenging examples. Furthermore, we explain how the technique may be applied to handle multiple option maturities, in a similar fashion as the method of \citet{andreasen2011volatility}. After writing this paper, we found out that the idea of using a piecewise-linear parametrization in the local variance gamma model was also developed in \citep{carr2018expanded, carr2019geometric}, with an additional stochastic drift term. As our focus is on building an arbitrage-free interpolation of market option quotes, and not on using the local variance gamma model to simulate the value of financial derivatives, we do not add a stochastic drift, and as a consequence, the equations we present here are much simpler. This allows for a significantly faster calibration.

We then propose a good regularization to calibrate the model against noisy option quotes. Finally, we put in evidence a flaw of the local variance gamma model, shared by the one-step finite difference technique of Andreasen and Huge, when the local variance function is at least of class $\mathcal{C}^1$ at the money, and propose two solutions for it.

\section{Dupire's PDDE in the local variance gamma model}\label{sec:pdde_solution}
We recall Dupire's partial difference differential equation (PDDE) for a call option price $C(T,x)$ of strike $x$ and maturity $T$ \citep{carr2017local}:
\begin{equation}
\frac{C(T,x)-\max(X(0)-x,0)}{T} =  \frac{1}{2}a^2(x)  \frac{\partial^2 C(T,x) }{\partial x^2} \,,\label{eqn:dupire_pdde}
\end{equation}
for a Martingale asset price process $X(t)$ of expectation $\mathbb{E_\mathbb{Q}}[X(t)] = X(0)$.

Let $\{ x_0, x_1,...,x_m, x_{m+1} \}$  be a increasing set of the strike prices, such that $x_0=L$, $x_{m+1}=U$ with the interval $(L,U)$ being the spatial interval where the  asset $X$ lives. Furthermore, we require the following to hold
\begin{equation*}\exists s \in \lbrack 1,m \rbrack | x_s = X(0)\,.\end{equation*}   
In practice, $(x_1,...,x_m)$ will correspond to the strike prices of the options of maturity $T$ we want to calibrate against, along with the forward price.

We consider $a$ to be a continuous piecewise-linear function with values $(\alpha_i)_{i=0,...,m}$ at the knots $(x_0,...,x_m)$. The local variance gamma volatility function reads
\begin{align}
a(x) &= \frac{\alpha_{i+1}-\alpha_i}{x_{i+1}-x_i} (x-x_i) + \alpha_i \quad \textmd{ for } x_i \leq x < x_{i+1}\,,\quad i=0,...,m\,.\ 
\end{align}

Let $V$ be the function defined by $V(x) = C(x,T)- \max(X(0)-x,0)$. $V$ is effectively the price of an out-of-the-money option (the price of a call option for $x > X(0)$ and of a put option for $x < X(0)$). The Dupire PDDE leads to
\begin{equation}
V(x) =  \frac{1}{2}a^2(x) T  V''(x) \,,\label{eqn:otm_ode}
\end{equation}
on the intervals $(L, X(0))$ and $(X(0),U)$. the function $V$ is not a solution of the above equation on the whole interval $(L, U)$ since $V'(x)$ jumps at $x=X(0)$. Indeed, the continuity of $\frac{\partial C}{\partial x}$ at $x=X(0)$ implies 
\begin{equation}\lim\limits_{x \to X(0)-} V'(x) = 1 + \lim\limits_{x \to X(0)^+} V'(x)\,.\label{eqn:jump_cond_v}\end{equation}
In order to define a unique $V$, we also impose the absorbing boundary conditions
\begin{equation}
V(L) = 0 = V(U)\,.\label{eqn:boundary_v}
\end{equation}
\section{Explicit solution}
When $\alpha_{i+1}=\alpha_i$, the  local variance gamma functional $a$ is constant on the interval $[x_i,x_{i+1}]$. The solution to Equation \ref{eqn:otm_ode} on this interval is given in \citet{carr2017local} and reads
\begin{align}
V(x) &= \chi_i(x) \left[\Theta_i^c \cosh\left(\omega_i(z_i(x)-z_i(x_i))\right) + \Theta_i^s \sinh\left(\omega_i(z_i(x)-z_i(x_i))\right)\right]\,,\label{eqn:lvg_V} 
\end{align}
with
\begin{equation*}
z_i(x) = x\,,\quad \omega_i = \frac{1}{\alpha_i}\sqrt{\frac{2}{T}}\,,\quad \chi_i =1\,.
\end{equation*}
We express the price in terms of hyperbolic functions, as it leads to a higher numerical stability, when compared to the exponential based expression of \citet{carr2017local}. The expression of $V$ on each interval is similar to a tension spline and we may use the fast and stable algorithm of \citet{renka1993algorithm} for the functions $\coshm(x) = \cosh(x)-1$ and $\sinhm(x)=\sinh(x)-x$, which gracefully handles the case when the argument $|x| < 0.5$. 

When $\alpha_{i+1} \neq \alpha_i$, the solution is of the same form, but 
with
\begin{equation*}z_i(x)= \ln |x + \frac{r_i}{q_i}|\,,\quad  \omega_i =\frac{1}{2} \sqrt{1+ \frac{8}{q_i^2 T}} \,,\quad \chi_i =\sqrt{\frac{x+\frac{r_i}{q_i}}{x_i+\frac{r_i}{q_i}}}\,.
\end{equation*}
with $q_i = \frac{\alpha_{i+1}-\alpha_i}{x_{i+1}-x_i}$, $r_i =\alpha_i - q_i x_i$.
The absolute value in $z_i(x)$ handles the case where the slope $q_i$ is negative.

\begin{proof}
	On the interval $[x_i,x_{i+1}]$, we have $a(x) = q_i x + r_i = \frac{\alpha_{i+1}-\alpha_i}{x_{i+1}-x_i} (x-x_i)+\alpha_i$. In particular, $a(x)$ will take values in the interval $[\alpha_i,\alpha_{i+1}]$. As $\alpha_i$ and $\alpha_{i+1}$ are assumed positive, $a(x)$ is also positive. 

	When the slope $q_i$ is positive, $x + \frac{r_i}{q_i}= \frac{a(x)}{q_i}$ is positive.  We may apply the change of variable $z=\ln(x+r_i/q_i)$, to Equation \ref{eqn:otm_ode}. This then leads to the following ODE
	\begin{equation*}
	V(z) = \frac{1}{2} q_i^2 T \frac{\partial^2 V}{\partial z^2} - \frac{1}{2} q_i^2 T \frac{\partial V}{\partial z} \,.
	\end{equation*}
	This is a second-order linear ODE with constant coefficients and the general solution is known to be of the form
	\begin{equation*}
	V(z) =  A_i e^{\lambda_i^{+} (z(x)-z(x_i))} + B_i e^{\lambda_i^{-} (z(x)-z(x_i))}\,,
	\end{equation*}
	where $\lambda_i^{+}$, $\lambda_i^{-}$ are the roots of the quadratic $\lambda^2 - \lambda - \frac{2T}{q_i^2} = 0$:
	\begin{equation*}
	\lambda_i^+ = \frac{1+\sqrt{1+ \frac{8}{q_i^2 T}}}{2}\,, \quad
	\lambda_i^- = \frac{1-\sqrt{1+ \frac{8}{q_i^2 T}}}{2}\,.
	\end{equation*}

	 When $q_i$ is negative, $-x-r_i/q_i = \frac{a(x)}{-q_i}$ is positive and we may apply the change of variable $z(x)=\ln(-x-r_i/q_i)$. This then leads to the same ODE as above.
	 
	The two cases may be merged together by defining transform $z(x) = \ln \left|x + \frac{r_i}{q_i}\right|$, with $z'(x)=\frac{1}{x + \frac{r_i}{q_i}}$.
\end{proof}

The derivative of $V$ reads
\begin{align}
V'(x) &= \chi_i(x) z_i'(x) \left[(\kappa_i\Theta_i^c + \omega_i\Theta_i^s) \cosh\left(\omega_i(z_i(x)-z_i(x_i))\right) + (\kappa_i\Theta_i^s + \omega_i\Theta_i^c) \sinh\left(\omega_i(z_i(x)-z_i(x_i))\right)\right]\,, 
\end{align}
with
\begin{align}
\kappa_i &=  \begin{cases}
0\quad \textmd{ when }  \alpha_i = \alpha_{i+1}\,,\\
\frac{1}{2} \quad \textmd {otherwise}\,,
\end{cases}\\
z_i'(x) &= \begin{cases}
1\quad \textmd{ when } \alpha_i = \alpha_{i+1}\,,\\
\frac{1}{x+\frac{r_i}{q_i}}\quad \textmd{ otherwise}\,.
\end{cases}
\end{align}

The boundary condition at $x=x_0=L$ translates to $\theta_0^c = 0$. One may choose $\theta_0^s$ arbitrarily to start the algorithm. and its relation with $V'$ is $\theta_0^s= \frac{V'(L)}{\omega_0}$. The final value will result from the continuity conditions at $i=s$.

The conditions to impose continuity of $V$ and its derivative at $x=x_{i+1}$ results in the following linear system
\begin{align}
\cosh_i \Theta_{i}^c +  \sinh_i \Theta_{i}^s &= \frac{\Theta_{i+1}^c}{\chi_i(x_{i+1})}\,, \label{eqn:price_cont}\\
(\kappa_i \cosh_i + \omega_i \sinh_i)  \Theta_i^c  + (\omega_i  \cosh_i + \kappa_i \sinh_i)\Theta_i^s &= 
\frac{\left(\kappa_{i+1}\Theta_{i+1}^c+ \omega_{i+1}\Theta_{i+1}^s\right)z_{i+1}'(x_{i+1})}{\chi_i(x_{i+1})z_i'(x_{i+1})} \label{eqn:der_cont}
\end{align}
for $i=0,...,s-2$,
with \begin{align*}
\cosh_i = \cosh\left(\omega_i(z_i(x_{i+1})-z_i(x_i))\right)\,,&\quad \sinh_i = \sinh\left(\omega_i(z_i(x_{i+1})-z_i(x_i))\right)\,.
\end{align*}
This system is solved from $i=0$, starting with the calculation of $\theta_{i+1}^c$ as given by Equation \ref{eqn:price_cont}. The value $\theta_{i+1}^c$ is then used to compute $\theta_{i+1}^s$ through Equation \ref{eqn:der_cont}.

At $x=x_{m+1}=U$, the boundary condition translates to 
\begin{align*}
\theta_{m}^s = V'(U) \frac{\cosh_m}{\omega_m z_m'(x_{m+1})}\,,&\quad \theta_{m}^c = -\theta_{m}^s \frac{\sinh_m}{\cosh_m}\,.
\end{align*}
Similarly to the lower boundary, we may start with an arbitrary $\theta_m^s$ and then solve Equations \ref{eqn:price_cont}, \ref{eqn:der_cont}, downwards from $i+1$ to $i$, for $i=m-1,m-2,...,s$.
. When $\cosh_i$ and $\sinh_i$ are very large (for example, in the case of a small slope $q_i$), the system may be ill-defined numerically and extra care must be taken. In this case, a simple solution is to solve approximately the system, by adding a small regularization constant: when $\frac{\sinh_i}{\cosh_i}=\pm 1$ numerically, we replace the latter ratio by $\pm 1 \mp \epsilon$.

The free parameters $\theta_0^s$ and $\theta_m^s$ are determined by the jump condition at $i=s$ defined by $V_{s-1}(x_s) = V_s(x_s)$ and $V_{s-1}'(x_s) = 1+V_s'(x_s)$. If $\theta^s_0$ is multiplied by a factor $\rho_L$, the system above implies that $\theta^c_i, \theta^s_i$ will also be multiplied by $\rho_L$ for $i<s$, and as a consequence, $V_{s-1}$ will be multiplied by $\rho_L$. Similarly, if $\theta^s_m$ is multiplied by a factor $\rho_R$, $V_s$ will be multiplied by $\rho_R$. We want to adjust $\theta^s_0$ and $\theta^s_m$ by so that $\theta^s_0 \rho_L$ and $\theta^s_m \rho_R$ verify exactly the jump condition. This leads to the system
\begin{align*}
V_{s-1}(x_s)\rho_L &= V_s(x_s)\rho_R\,,\\
V_{s-1}'(x_s)\rho_L &= 1+ V_s'(x_s) \rho_R\,,
\end{align*}
with 
\begin{align*}
V_{s-1}(x_s) &= \chi_{s-1}(x_s) (\theta_{s-1}^c \cosh_{s-1} + \theta_{s-1}^s \sinh_{s-1}) \,,\\
V_s(x_s) &= \theta_s^c\,,\\
	V_{s-1}'(x_s) &=\chi_{s-1}(x_s) z_{s-1}'(x_s) \left[ (\kappa_{s-1} \theta_{s-1}^c + \omega_{s-1} \theta_{s-1}^s)\cosh_{s-1} +  (\omega_{s-1} \theta_{s-1}^c + \kappa_{s-1} \theta_{s-1}^s) \sinh_{s-1}\right] \,,\\
	V_s'(x_s) &= (\kappa_s\theta_s^c + \omega_s\theta_s^s) z_s'(x_s)\,.
\end{align*}
And the solution is 
\begin{align}
\rho_R &= \frac{1}{\frac{V_s(x_s)}{V_{s-1}(x_s)}V_{s-1}'(x_s) - V_s'(x_s)}\,,\label{eqn:rhor}\\
\rho_L &= \frac{V_s(x_s)}{V_{s-1}(x_s)}\rho_R\,.\label{eqn:rhol}
\end{align}
In order to make the solution verify the jump condition, we then multiply $(\theta^c_i)_{i=0,...,s-1}$, $(\theta^s_i)_{i=0,...,s-1}$ by $\rho_L$, and $(\theta^c_i)_{i=s,...,m}$, $(\theta^s_i)_{i=s,...,m}$  by $\rho_R$.

So far, we have ensured that $V$ is a solution to Equation \ref{eqn:otm_ode} of class $\mathcal{C}^1$ on the intervals $(L,X(0))$ and $(X(0),U)$. By definition, $C(x,T) = V(x) + \max(X(0)-x,0)$. For $i \in \{1,...,s-1,s+1,...,m\}$, we have $V_i''(x_i) = 2 a^2(x_i) V_i(x_i)$, and $V_{i-1}''(x_i) =  2 a^2(x_i) V_{i-1}(x_i)$. By construction, we know that $V_i(x_i) = V_{i-1}(x_{i})$ and $a$ is continuous, thus $V''$ is continuous at $x_i$, and we deduce that $C \in \mathcal{C}^2(L,X(0))$, $C \in \mathcal{C}^2(X(0),U)$. At $x=X(0)$, the jump condition and the continuity of $a$ ensures that $\frac{\partial^2 C}{\partial x^2}$ is also continuous. We have thus obtained a solution of class $\mathcal{C}^2$  to the Dupire PDDE.

\section{Calibration}
\subsection{Single maturity}\label{sec:calibration_single}
The calibration of a single maturity consists in finding the parameters $\alpha_0, ..., \alpha_{m+1}$ such that the market option prices $(\hat{C}_i)_{i=1,...m}$ of respective strikes $(x_i)_{i=1,...,m}$ match exactly the function $C(x,T)$, solution of the Dupire PDDE of class $\mathcal{C}^2$, at each market option strike.

The problem is under-determined as we have $m+2$ parameters and $m$ reference prices. In order to resolve this discrepancy, the parameters $\alpha_0$ and $\alpha_{m+1}$ may be set arbitrarily to control the extrapolation. We choose a flat extrapolation in terms of the local variance gamma function $a$, that is, $\alpha_0 = \alpha_1$ and $\alpha_{m+1}=\alpha_{m}$. 

In general, the market strikes will not include $X(0)$. In this case, $X(0)$ must be added to the knots $\{x_i\}_{i=1,...,m}$ used in the local variance gamma representation. This adds one more parameter $\alpha_s$ to the representation, where $s$ is the index corresponding to $X(0)$ in the set of knots. We take this parameter to be the linear interpolation of the enclosing parameters: $\alpha_s = \frac{x_{s}-x_{s-1}}{x_{s+1}-x_{s-1}} \sigma_{s+1} + \frac{x_{s+1}-x_{s}}{x_{s+1}-x_{s-1}}  \sigma_{s-1}$.

The calibration problem may be expressed as the least-squares minimization of the error measure $E$ defined by
\begin{equation}\label{eqn:objective}
E = \sum_{i=1}^m \mu_i^2\left(\sigma(\alpha, x_i) - \hat{\sigma}_i\right)^2\,,
\end{equation}
with $\alpha_i > 0$ for $i=1,...,m$ and where $\sigma(\alpha, x)$ is the implied volatility corresponding to the option prices obtained with the piecewise-linear local gamma variance model and $\hat{\sigma}_i$ is the market implied volatility at strike $x_i$, $(\mu_i)_{i=1,...,m}$ are weights associated to the accuracy of the fit at each point. 

In order to solve this non-linear least-squares problem, we will use the Levenberg-Marquardt algorithm as implemented by \citet{klare2013gn}. The box constraints $\alpha_i > 0$ can be added in a relatively straightforward manner to any Levenberg-Marquardt algorithm, through the projection technique described in \citep{kanzow2004levenberg}, or through a variable transform from $\mathbb{R}$ to a subset of $\mathbb{R^+}$ (for example through the function $x \to x^2 + \epsilon$ with some small positive $\epsilon$). 

The implied volatility for a given option price may be found efficiently and accurately through the algorithm of \citet{jackel2013let}. Alternatively, we may directly solve an almost equivalent formulation in terms of option prices, using the error measure $E_V$ defined by
\begin{equation}\label{eqn:objective_price}
E_V = \sum_{i=1}^m w_i^2\left(C(\alpha, x_i) - \hat{C}_i\right)^2\,,
\end{equation}
with $C(\alpha, x)$ being the local variance gamma option price with parameter $\alpha$ and strike $x$, and the capped inverse Vega weights $w_i$ given by
\begin{equation}
w_i = \min\left(\frac{1}{\nu_i}, \frac{10^6}{X(0)} \right)\mu_i\,,
\end{equation}
where $\nu_i= \frac{\partial \hat{C}_i}{\partial \sigma}$ is the Black-Scholes Vega corresponding the market option price $\hat{C}_i$, and $10^6$ is a cap applied to avoid numerical issues related to the limited machine accuracy (see Appendix \ref{sec:weights} for an explanation of this relation between $w$ and $\mu$).

\subsection{Alternative calibration strategies}
Another approach would be to use a fixed-point method, similar to the one described in \cite{reghai2006hybrid,reghai2012local}. 

One could also explore solving exactly each $\alpha_i$ successively using a one-dimensional non-linear solver, given an initial guess for $A_0$, and then either solve for $A_0$, or use a fixed point iteration on $A_0$. Each iteration would require to solve again the $(\alpha_i)_{i=1,...,m}$.

We found the performance of a straightforward Levenberg-Marquardt minimization acceptable in practice, and the least-squares approach to be more flexible if some additional regularization is needed. As initial guess, we found that various simple choices were almost equally effective such as: $\sigma_{\textsf{atm}} X(0)$, or $\sigma_{\textsf{atm}} x_i$, corresponding to an approximately flat Bachelier guess or an approximately flat lognormal guess, with  $\sigma_{\textsf{atm}}$ being the (approximate) at-the-money Black-Scholes implied volatility in the set of options to fit.

\subsection{Multiple maturities}\label{sec:calibration_multiple}
In the real world, the quoted options are likely to be on an asset with a non-zero time-dependent drift. For example, an equity will involve the interest rate and dividend yield evolution up to the maturity of the option, and a foreign exchange rate will involve the domestic and foreign interest rates evolution.
As described in \citep{buehler2010volatility}, it is always possible to translate the problem towards a problem on a driftless process $X$, with call prices on a scaled strike. For example, for an asset $S$ with forward price to time $t$, $F(0,t)=\mathbb{E}[S(t)]$, we may consider $X(t) = \frac{S(t)}{F(0,t)}$. The process $X$ is a martingale and $\mathbb{E}[X(t)]=1$. Then, the price $C$ of a call option of maturity $T_1$ and strike $x$ on $X$ is related to the price $\mathbb{C}$ of a call option of maturity $T_1$ on $S$ by
\begin{align}C(T_1,x) &= \frac{1}{B(0,T_1) F(0,T_1)} \mathbb{C}(T_1, F(0,T_1)x)\,,\end{align}
and thus the problem with strikes $K_i$ on the asset $S$ is equivalent to a problem with strikes $x_i=\frac{K_i}{F(0,T_1)}$ on $X$. If a second maturity $T_2$ is involved, with the same strikes $K_i$, the original quotes will be converted to quotes on $X$ at different set of strikes $x_i=\frac{K_i}{F(0,T_2)}$.

Let us consider two option maturities $T_1, T_2$ with $T_2 > T_1 > 0$. Let $\mathcal{x}_1 = \left\{x_{1,1},...,x_{1,m_1}\right\}$ be the set of strikes for the first maturity and $\mathcal{x}_2 =\left\{x_{2,1},...,x_{2,m_2}\right\}$ the strikes for the second maturity.
We start by calibrating the first maturity as described in Section \ref{sec:calibration_single} using the knots $\left\{L,X(0),U\right\} \cup \mathcal{x}_1$. This leads to the parameters $(\alpha_{1,i})$. Then we add $\mathcal{x}_1 - \mathcal{x}_1 \cap \mathcal{x}_2$ to the knots for the second maturity, and use a linear interpolation to define the extra parameters $\alpha_{2,i}$ corresponding to the set $\mathcal{x}_1 - \mathcal{x}_1 \cap \mathcal{x}_2$.

We would like then to solve Dupire's PDDE from $T_1$ to $T_2$:
\begin{equation}
\frac{C(T_2,x) - C(T_1,x)}{T_2-T_1} = \frac{1}{2}a^2(x) \frac{\partial^2 C(T_2,x)}{\partial x^2}\,,\label{eqn:dupire_pdde_2mat}
\end{equation}
If $C(T_1,x)$ is obtained from a calibrated LLVG model between 0 and $T_1$ and is thus given by Equation \ref{eqn:lvg_V}, then Equation \ref{eqn:dupire_pdde_2mat} dictates that $C(T_2,x)$ will not have the same analytical form, unless the local variance function $a(x)$ is the same at both time-steps, or is a piecewise-constant function. We would however like to find a piecewise-linear function $a(x)$ at $T_2$, different from the one used to calibrate the options of maturity $T_1$, in order to fit the quotes at $T_2$, and then it is not obvious if there is an analytical solution at all.

Instead of using a calibrated LLVG model for $C(T_1, x)$, we may consider a continuous piecewise linear interpolation of the LLVG option prices at $T_1$. Typically we would use the market strikes as knots of this interpolation, but we may also use a finer grained representation, based on a calibrated LLVG model at $T_1$. Then, $C(T_2, x)$ will be of the form given by Equation \ref{eqn:lvg_V}. The solution for $\theta^c, \theta^s$ is however slightly different, since the first derivative must jump at each knot, instead of only at $X(0)$, but the same technique is fully applicable. 
It is then guaranteed that at each knot of the piecewise linear interpolation, $C(T_2, x) > C(T_1, x)$, which means that there is no calendar spread arbitrage at those points. It is not guaranteed everywhere, but we can always increase the number of knots if there is a concern.


A much more straightforward approach, which practitioners sometimes apply, would be to calibrate the two maturities independently. As the calibration is almost exact, and option quotes are arbitrage-free, we know that this approach leads to no arbitrage at the discrete set of points corresponding to the union of all the options we use in the calibration. Of course, there may however be some spurious calendar spread arbitrage at a strike in between two quotes strikes. The hope is that it is a rare occurrence and we may find other ways to deal with those cases then.

Clearly, this is an area where the technique of \citet{andreasen2011volatility} has an edge over the LLVG model, as the former will guarantee the absence of  calendar spread arbitrages on a denser grid, and allows for any function $a(x)$ (not only piecewise-constant or piecewise-linear).


\subsection{Dealing with arbitrages in the input quotes}
The input market prices, or equivalently, their implied volatilities, may not be arbitrage-free. In this case, the  local variance $\alpha_i$ will either move towards zero (in case of a calendar spread arbitrage) or towards infinity (in case of a butterfly spread arbitrage) and the minimization will not be exact. It will be thus important to cap and floor the local variance during the minimization. 

An alternative approach is to de-arbitrage the input quotes via the quadratic programming representation described in Appendix \ref{sec:arbitrage_free_qp}. The quadratic programming problem is fast to solve using a standard optimization library such as CVXOPT \citep{andersen2013cvxopt}, OSQP \citep{osqp} or quadprog \citep{turlach2007quadprog}, typically much faster than the calibration. A side-effect of this method is to remove any outlier automatically: they will effectively be smoothed out by the de-arbitraging process. In the direct approach described previously, outliers may skew significantly the result, although one may remedy this by assigning smaller weights to the identified outliers, which may be related to a large bid-ask spread. For example, an inverse bid-ask spread weighting scheme may be effective.

Even if the quotes are de-arbitraged, they may still be noisy. As a consequence, the implied probability density would present many spikes, and the option gamma sensitivity may be of poor quality.

In those cases, it may be useful to add regularization to the minimization as well.
An interesting candidate for the regularization is to minimize the strain energy of the beam that is forced to pass through the given data points \citep{glass1966smooth}:
\begin{equation*}\label{eqn:objective_reg_beam}
E = \sum_{i=1}^m \mu_i^2\left(\sigma(\alpha, x_i) - \sigma_i\right)^2 +\lambda^2 \sum_{i=1}^{m-1}  \frac{\mu_i^2 \sigma''(\alpha, x_i)}{\left[1+\sigma'(\alpha, x_i)^2\right]^{\frac{5}{2}}}(x_{i+1}-x_i)\,,
\end{equation*}
The first term of the objective $E$ corresponds to the square of the RMSE, while the second term is the regularization. The regularization parameter $\lambda$ controls the smoothness of the spline interpolation. 
We may wish to use a simpler version, where we regularize directly the local variance function, by penalizing differences in the three-points estimate of its second derivative:
\begin{equation}\label{eqn:objective_reg_alpha}
E = \sum_{i=1}^m \mu_i^2\left(\sigma(\alpha, x_i) - \sigma_i\right)^2 +\tilde{\lambda}^2 \sum_{i=2}^{m-1}  \frac{ \alpha_{i+1}-\alpha_i}{x_{i+1}-x_{i}} - \frac{  \alpha_{i}-\alpha_{i-1}}{x_{i}-x_{i-1}}\,,
\end{equation}
with $\tilde{\lambda}^2 =  2 X(0)^2  \lambda^2 \sum_{i=1}^m \mu_i^2$. We will see however that, while it leads to a smooth density, it also results in some spurious spike in the density, located at-the-money, at $x=X(0)$.
Our goal is to obtain a smooth probability density, and we may directly perform the regularization on the density via
\begin{equation}\label{eqn:objective_reg_density}
E = \sum_{i=1}^m \mu_i^2\left(\sigma(\alpha, x_i) - \sigma_i\right)^2 +\tilde{\lambda}^2 \sum_{i=2}^{m-1}  \frac{ \ln V''(x_{i+1})-\ln V''(x_i)}{x_{i+1}-x_{i}} - \frac{ \ln V''(x_{i})-\ln V''(x_i-1)}{x_{i}-x_{i-1}}\,,
\end{equation}
where $V''(x_i) = \frac{2\theta_i^c}{a(x_i)^2} =\frac{2\theta_i^c}{\alpha_i^2}  $. 
\section{Numerical examples}
\subsection{Exact interpolation without wiggles}
\citet{jackel2014clamping} shows that undesired oscillations  can appear  in the graph of the implied volatility against the option strikes when the option prices are interpolated by a monotonic and convex spline. Table \ref{tbl:jackel_clamping_1} in appendix \ref{sec:market_data_jackel} presents a concrete example\footnote{We are grateful to Peter J\"ackel for kindly providing this data.}. Here, the option quotes are not direct market quotes, but the solution of a sparse finite difference discretization of a local stochastic volatility model: the market never quotes so far out-of-the-money option prices. His data has a few interesting properties: 
\begin{itemize}
	\item some of the option prices are extremely small: the interpolation must be very accurate numerically.
	\item the option prices are free of arbitrage. In theory, an arbitrage-free interpolation can be exact.
	\item a cubic spline interpolation on the volatilities or the variances, often used by practitioners, is not arbitrage-free.
	\item a convexity preserving $C^1$-quadratic, or $C^2$-rational spline results in strong oscillations in the implied volatility.
\end{itemize} 
The interpolation proposed in \citep{jackel2014clamping} possesses unnatural spikes at the points of clamping, in particular, the implied density is not continuous.

\begin{figure}[h]
	\centering{
		\subfigure[\label{fig:jaeckel_decf_vol}Volatility smile for log-moneyness $\log\frac{x}{X(0)}$ larger than zero.]{
			\includegraphics[width=.9\textwidth]{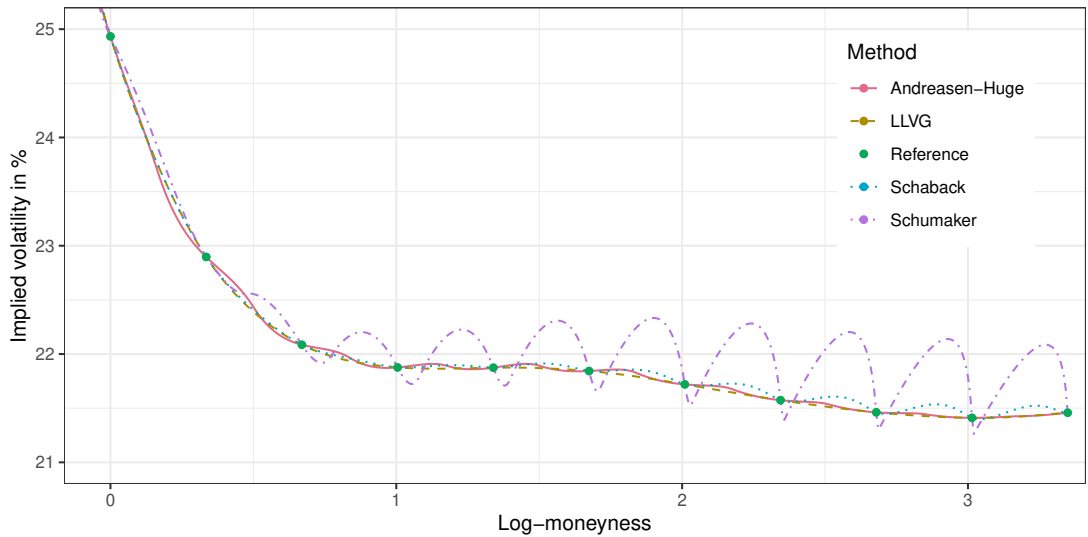}}
		\subfigure[\label{fig:jaeckel_decf_dens}Probability density.]{
			\includegraphics[width=.9\textwidth]{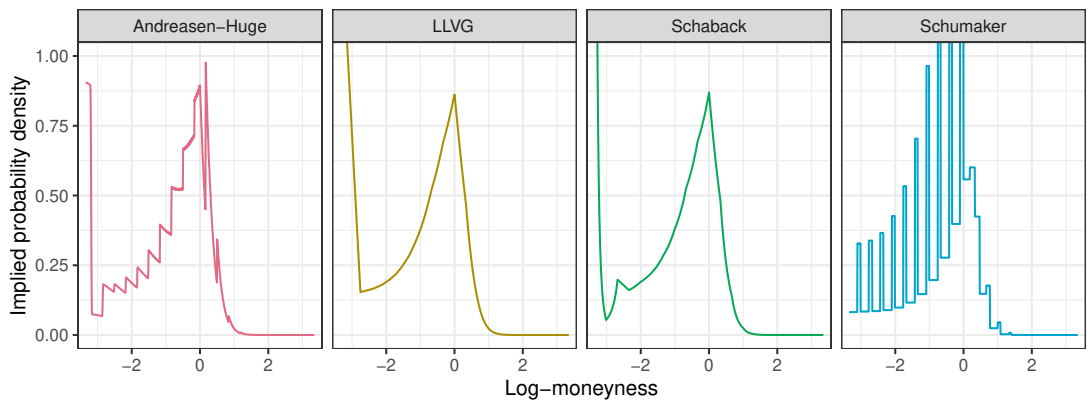} }
		\caption{Implied volatility and probability density for Case I (Table \ref{tbl:jackel_clamping_1} in appendix \ref{sec:market_data_jackel}).}}
\end{figure}

We apply the quadratic convex spline interpolation of \citet{schumaker1983shape}, the rational convex spline of \citet{schaback1973spezielle}, the one-step finite difference method of \cite{andreasen2011volatility}, and our linear local variance gamma (LLVG) interpolation on those quotes. As already evidenced in \citep{jackel2014clamping}, the convex spline interpolations of the option prices lead to unnatural oscillations in the implied volatility. The quadratic spline results in much stronger oscillations. The Andreasen-Huge technique is not exempt of oscillations, if a piecewise constant discrete local volatility representation is used, as in Figure \ref{fig:jaeckel_decf_dens}. 
The discontinuities in the representation leads to discontinuities in the implied probability density. If, instead of a piecewise constant function, we use a piecewise linear function, the oscillations in the implied volatilities with the technique of Andreasen and Huge disappear. The LLVG interpolation does not present any oscillations in the implied volatilities and results in a smooth implied probability density.

Furthermore, it is nearly exact on this example, and slightly faster than Andreasen and Huge technique applied on a grid of 400 points (Table \ref{tbl:jackel_collo_rmse}). It is however still more than an order of magnitude slower than the convex spline interpolations. 
\begin{table}[h]
	\caption{Root mean square error (RMSE) of the interpolated implied volatilities against the implied volatilities of Table \ref{tbl:jackel_clamping_1} in Appendix \ref{sec:market_data_jackel}. For Andreasen-Huge and LLVG, the solver error tolerance is set to $10^{-8}$.\label{tbl:jackel_collo_rmse}}
	\centering{
		\begin{tabular}{llrlr}\toprule
			Method & \multicolumn{2}{c}{Case I} & \multicolumn{2}{c}{Case II} \\
			& RMSE & Time (ms) & RMSE & Time (ms)\\\cmidrule(lr){2-3}\cmidrule(lr){4-5}
		LLVG &  $2\cdot10^{-13}$ &2.00 & $2\cdot10^{-8}$ &29.20 \\	
		Andreasen-Huge (400 nodes, flat) &   $4\cdot10^{-15}$ & 10.90 & $6\cdot10^{-4}$ & 11.70 \\
Andreasen-Huge (400 nodes, linear) &   $5\cdot10^{-12}$ & 3.20 & $2\cdot10^{-5}$ &34.60 \\
			Schaback convex spline  & $5\cdot10^{-16}$ & 0.01 & $8\cdot10^{-16}$ & 0.02\\
			Schumaker convex spline & $5\cdot10^{-16}$ & 0.02 & $8\cdot10^{-16}$ & 0.01\\\bottomrule
	\end{tabular}}
\end{table}

\begin{figure}[!h]
	\centering{
		\subfigure[\label{fig:jaeckel_inc_vol}Volatility smile for the LLVG model.]{
			\includegraphics[width=.9\textwidth]{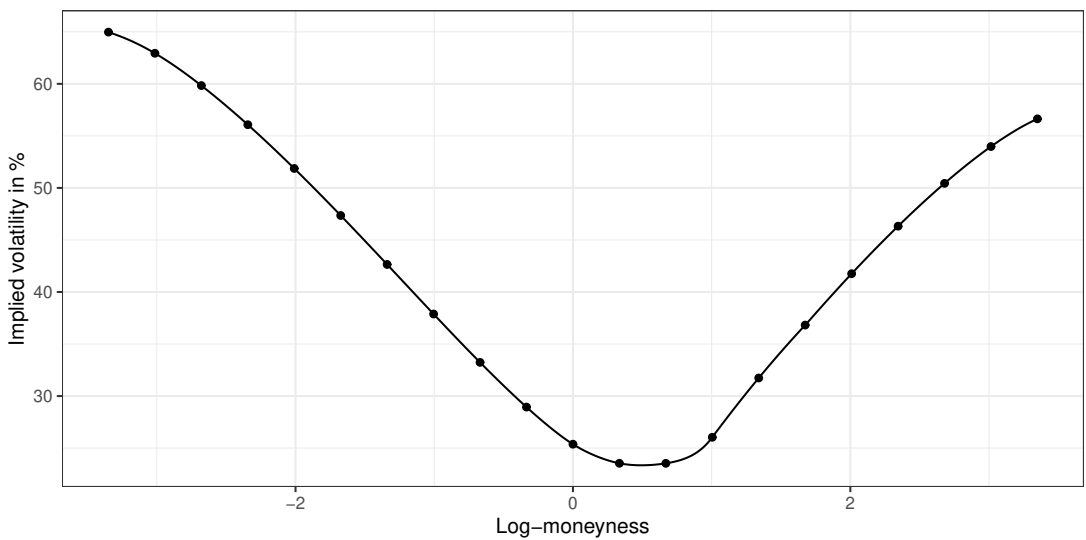}}
		\subfigure[\label{fig:jaeckel_inc_dens}Probability density in log-scale.]{
			\includegraphics[width=.9\textwidth]{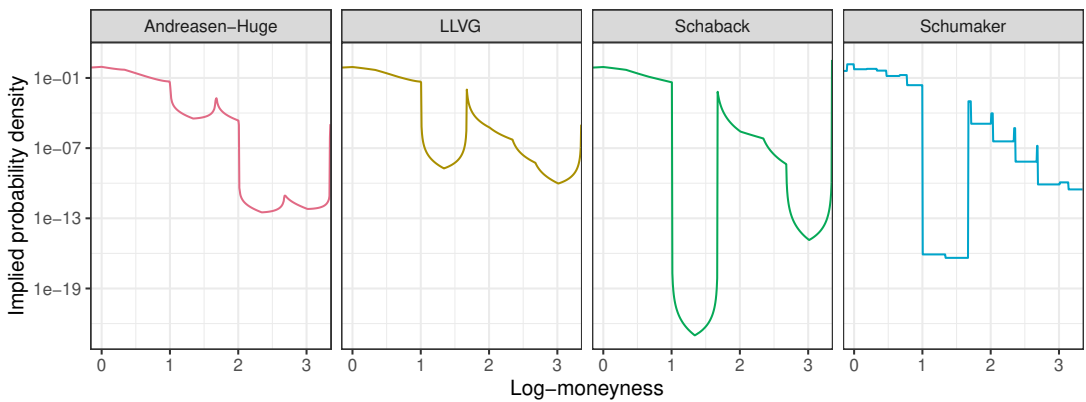} }
		\caption{Implied volatility and probability density for Case II (Table \ref{tbl:jackel_clamping_1} in appendix \ref{sec:market_data_jackel}).}}
\end{figure}
The second example (Case II of Table \ref{tbl:jackel_clamping_1}) is more challenging numerically, since some option prices are very close to an arbitrage: the difference of consecutive option price slopes $\frac{c_{i+1}-c_i}{x_{i+1}-x_i}-\frac{c_{i}-c_{i-1}}{x_{i}-x_{i-1}}$, at strike $x_i=3.81$, is around  $10^{-16}$, close to machine epsilon accuracy. As a consequence, the Levenberg-Marquardt solver takes more iterations to find the solution for the LLVG model, and finds only an approximate solution for Andreasen and Huge method (even on finer grids). All of the methods leads to a similar implied volatility interpolation, and there is no oscillation on this example (Figure \ref{fig:jaeckel_inc_vol}).

\begin{figure}[h]
	\centering{
		\includegraphics[width=.9\textwidth]{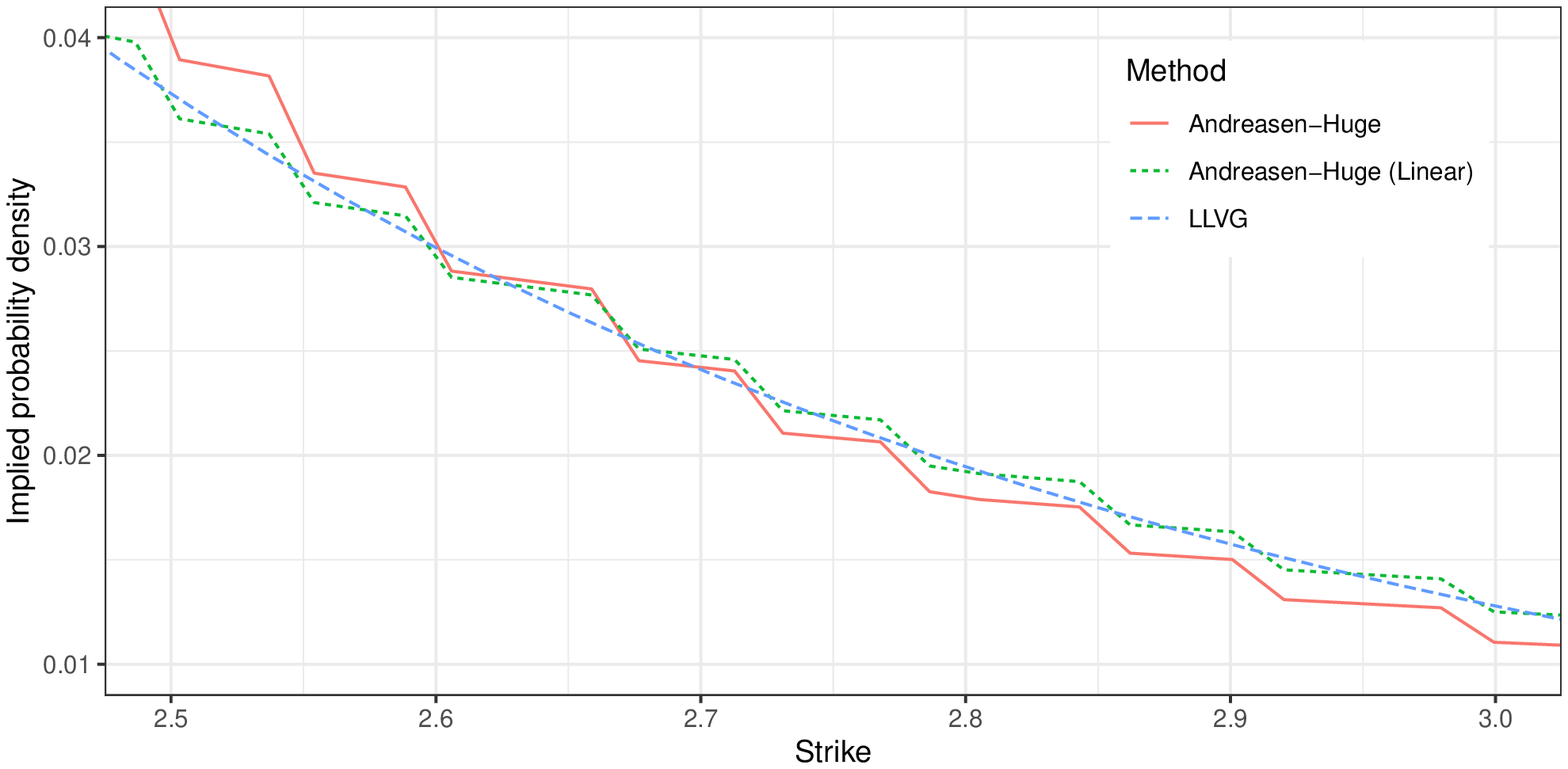}}
	\caption{Implied probability density for the LLVG and Andreasen-Huge methods, for strike moneyness $K \in [2.5, 3.0]$, calibrated to the market data of Table \ref{tbl:jackel_clamping_1} in Appendix \ref{sec:market_data_jackel}.\label{fig:jaeckel_wiggle_decf_density_zoom}}
\end{figure}

The optimal implied probability density corresponding to Case I is relatively smooth everywhere, and especially for log moneyness larger than zero. Figure \ref{fig:jaeckel_wiggle_decf_density_zoom} shows however that the probability density implied by technique of \citet{andreasen2011volatility} exhibits a staircase shape when zoomed-in, with both piecewise-constant and piecewise-linear representations. This is due to the interpolation in between the finite difference grid nodes. In contrast, the probability density implied by the LLVG model stays very smooth, and is continuous by construction.

\subsection{Interpolation of noisy data}
We consider option quotes on the SPX500 index of maturity 1 month, as of March 18, 2020, taken from the Chicago Board Options Exchange (CBOE). For each strike and maturity, at a given time, the market quotes two prices for an option contract: the bid price and the ask price.
In order to calibrate directly the LLVG model to the market quotes, we need a single estimate of the implied cumulative probability density.
It is common practice to use the average of the bid and ask prices, the mid price for this purpose. Alternatively, we could also build two distinct representations: one for the bid prices and one or the ask prices. 

When taken separately, the bid, ask or mid prices are not guaranteed to be arbitrage-free in theory: there can be theoretical arbitrages within the  bid-ask spread that can not be taken advantage of in practice.

\begin{figure}[!h]
		\subfigure[\label{fig:vol_spx_mar18_2020_1_lvg}Volatility smile for the LLVG model.]{
			\includegraphics[width=\textwidth]{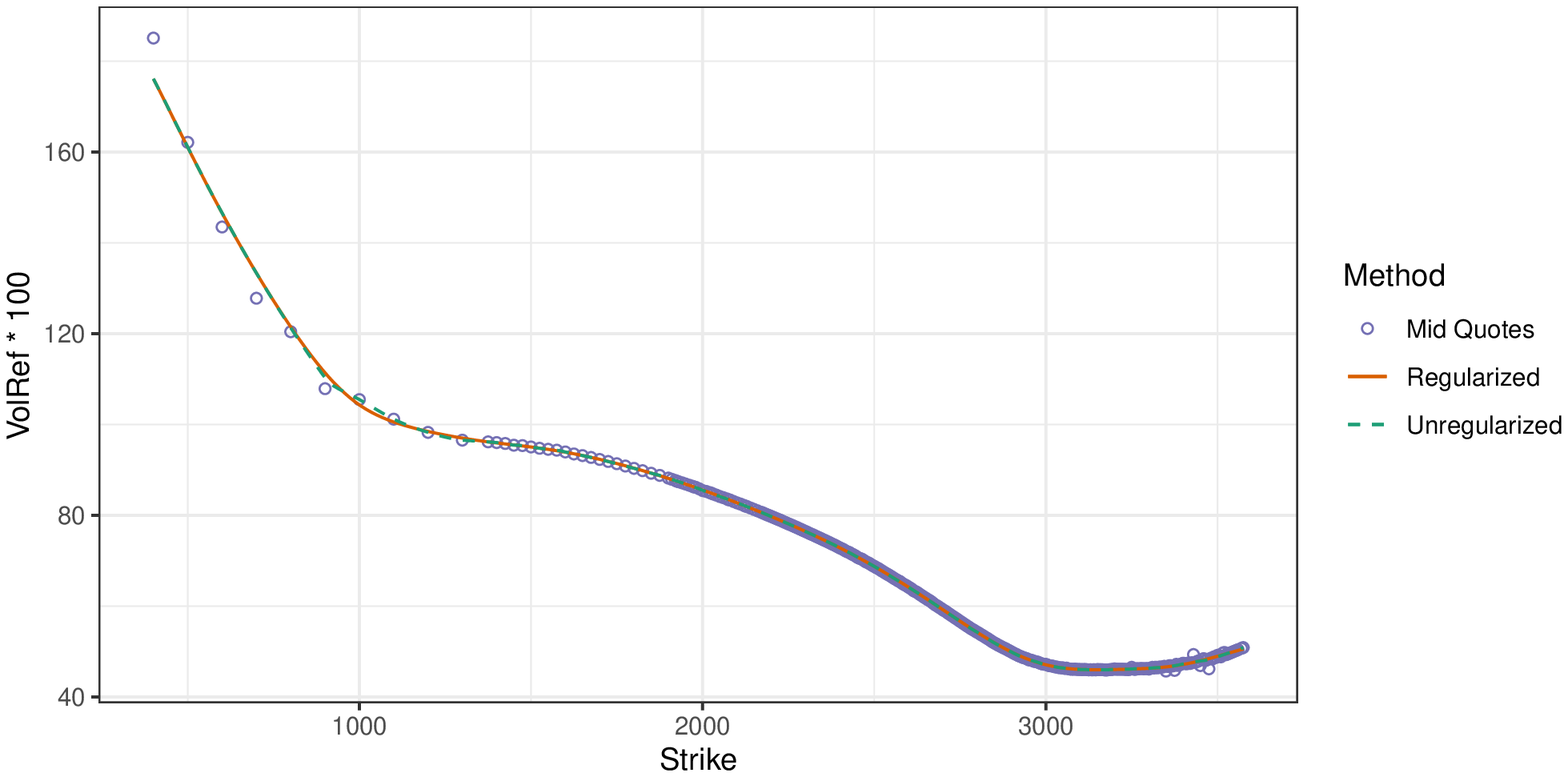}}
		\subfigure[\label{fig:density_spx_mar18_2020_1_lvg}Probability density.]{
			\includegraphics[width=\textwidth]{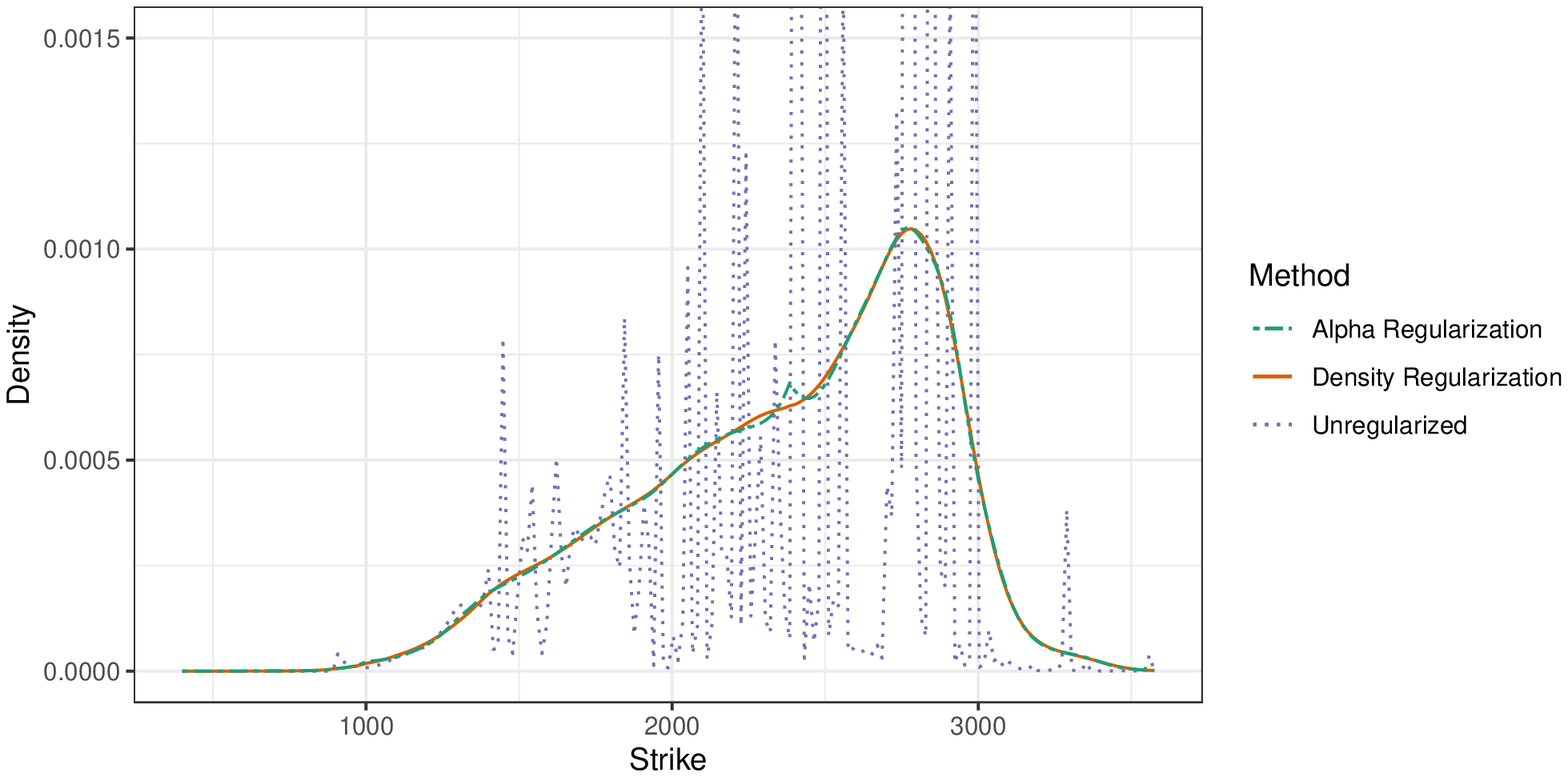} }
		\caption{Implied volatility and probability density of 1m SPX options as of March 18, 2020.}
\end{figure}

If we calibrate the LLVG model towards the closest arbitrage-free quotes, using the algorithm described in Appendix \ref{sec:arbitrage_free_qp}, the fit is not exact, mainly for two reasons: the problem is high-dimensional as there are 344 quotes, and the problem is not well conditioned, since the optimal implied density is extremely jagged as evidenced in Figure \ref{fig:density_spx_mar18_2020_1_lvg}.

The addition of a regularization during the calibration is thus a necessity. The regularization on the parameters $\alpha$ (Equation \ref{eqn:objective_reg_alpha}) is not appropriate here. It results in a spurious spike located at the forward price $X(0)=2384$. The regularization on the probability density (Equation \ref{eqn:objective_reg_density}) works well.

The reason for the appearance of a spike with the regularization on the parameters $\alpha$ is deeply linked to the local variance gamma model itself. Indeed, if we set the parameters to a constant, for example $\alpha=0.2$, the implied probability density exhibits a strong spike at $X(0)$. This is relatively intuitive from Equation \ref{eqn:dupire_pdde} since the left hand side $C(T,x)-\max(X(0)-x,0) $ is composed of a smooth function minus a function with a discontinuous derivative at $X(0)$.  

With noisy quotes, the regularization helps to  significantly reduce the calibration time as the problem is better conditioned (Table \ref{tbl:perf_spx_mar18_2020_1_lvg}). It takes around 1 second to calibrate the 344 option prices of maturity 1 month, and 0.20 second to calibrate the 165 option prices of maturity 2 months, using an Intel Core i7 7600U processor.
\begin{table}[h]
	\caption{Root mean square error (RMSE) of the interpolated implied volatilities against the implied volatilities of Table \ref{tbl:jackel_clamping_1} in Appendix \ref{sec:market_data_jackel}. \label{tbl:perf_spx_mar18_2020_1_lvg}} 
	\centering{
		\begin{tabular}{lrrrr}\toprule
			Method & \multicolumn{2}{c}{Case III} & \multicolumn{2}{c}{Case IV} \\
			& RMSE & Time (s) & RMSE & Time (s)\\\cmidrule(lr){2-3}\cmidrule(lr){4-5}
			LLVG regularized &  0.00137 &0.99  & 0.00027 & 0.20\\ 
			LLVG unregularized & 0.00126 &5.86 & 0.00002 & 4.00\\ \bottomrule
	\end{tabular}}
\end{table}

\subsection{The spurious spike at $X(0)$}
\begin{figure}[h]
	\centering{
		\includegraphics[width=.9\textwidth]{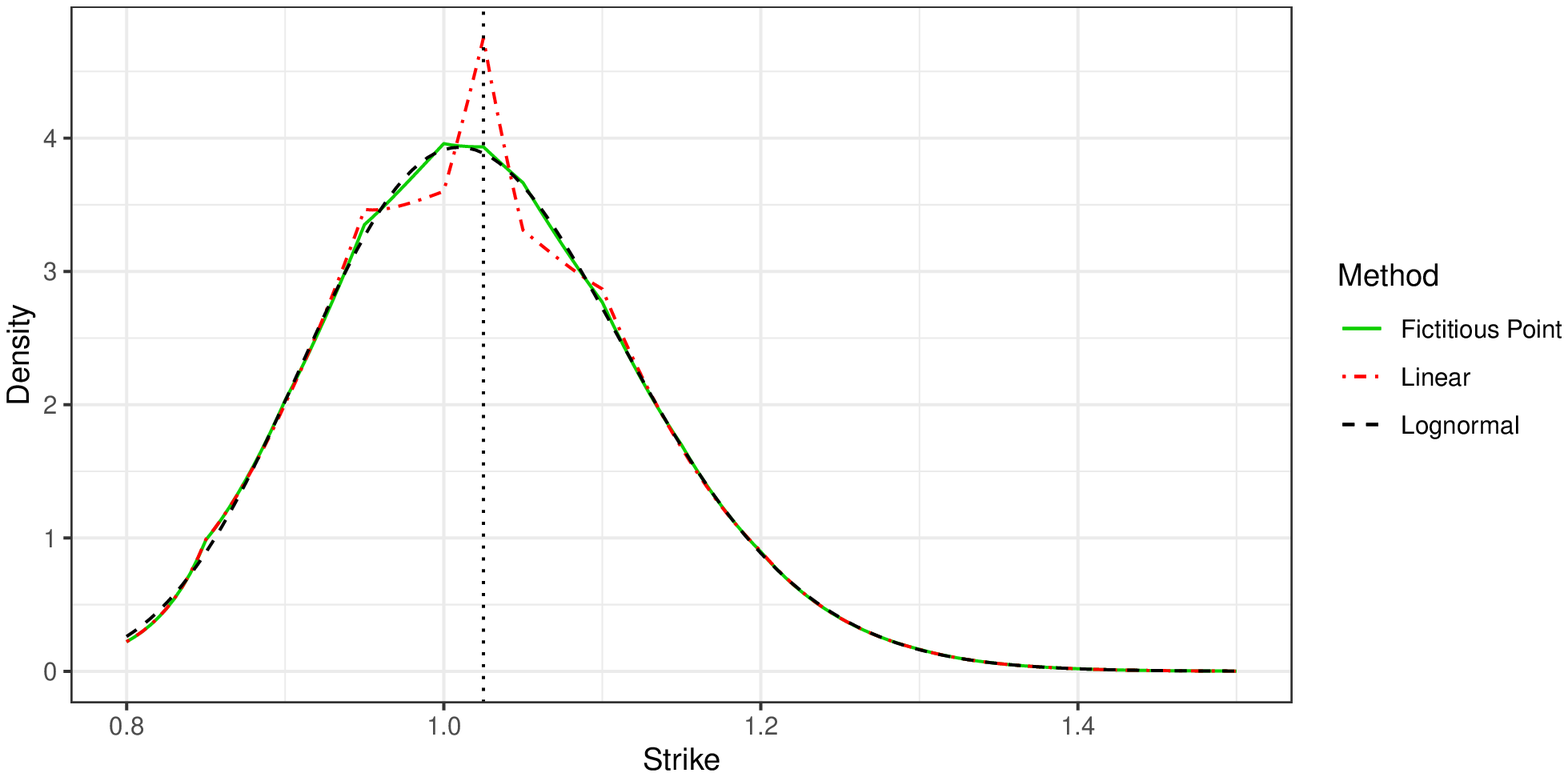}}
	\caption{Implied probability density for the LLVG model, using a fictitious point with regularization, or a linear interpolation for the parameter $\alpha_s$, fitted to a Black-Scholes model with constant volatility $\sigma_B=20\%$, time to expiry $T=0.25$ and forward $1.025$ (dotted vertical line).\label{fig:density_blackflat_lvg}}
\end{figure}

\begin{figure}[h]
	\centering{
		\includegraphics[width=.9\textwidth]{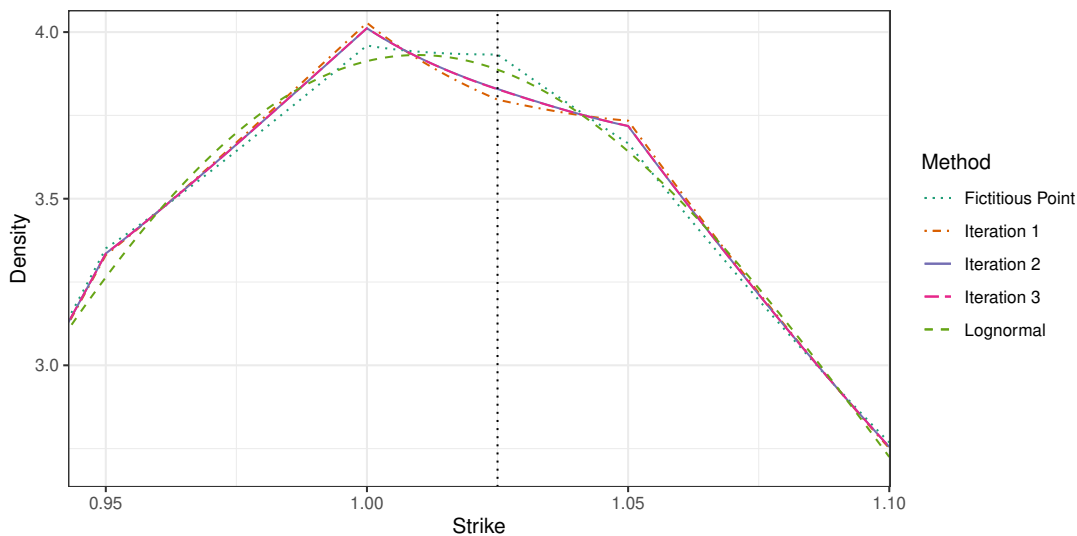}}
	\caption{Zoom of Figure \ref{fig:density_blackflat_lvg}, showing probability density implied by the calibration of the LLVG model with one, two, or three iterations to update $\sigma_s$.\label{fig:density_blackflat_lvg_zoom}}
\end{figure}

The regularization on the density may not always fully solve the issue: when the number of knots (which correspond to the number of quotes) is small, and the peak density is not close to the forward price, the spike may appear in between the two knots which encompass the forward price. The Andreasen and Huge technique suffers from the same issue. When fitting to market data, it is not easy to find an example, since, with a few points, the implied distribution will not tend to be very smooth in general. As an illustration, we thus consider a manufactured example: we fit the LLVG model to 10 option prices of strikes (0.85, 0.90, 0.95, 1, 1.05, 1.1, 1.15, 1.2, 1.3, 1.4), obtained by the Black-Scholes model with constant volatility $\sigma_B=20\%$, time to maturity $T=0.25$ and forward price $1.025$. We know that the theoretical distribution is a lognormal distribution. When $\alpha_s$ is computed by linear interpolation, a large spike appears in the probability density implied from a calibrated LLVG model, even with regularization (Figure \ref{fig:density_blackflat_lvg}). 
When we add a fictitious point at $x=1.025$ with a weight of zero ($\alpha_s$ is left as a free parameter), and use a regularization constant $\lambda=10^{-8}$ during the calibration, the spike disappears and the implied probability density is much closer to the lognormal distribution. On this example, the RMSE in implied volatilities is $3\cdot 10^{-7}$ with regularization, the fit is almost exact. An alternative which does not require regularization, is to choose $\alpha_s$ such that our piecewise representation $V$ is of class $\mathcal{C}^3$ in the interval $(x_{s-1},x_{s+1})$. By applying the derivative to $V''$, along with Equation \ref{eqn:otm_ode}, the $\mathcal{C}^3$ continuity relation at $x=x_s$ reads
\begin{equation*}
\left(\frac{V}{a^2}\right)' (x_s^-) = \left(\frac{V}{a^2}\right)' (x_s^+) \,,
\end{equation*}
where the notation $x_s^-$, $x_s^-$ denotes the value of the limit towards $x_s$ respectively from the left and from the right. Using the continuity of $V(x_s)=\theta^c_s$, the jump condition of $V'$ at $x_s$ and the continuity of $a(x_s)= \alpha_s$, this leads to
\begin{equation*}
\frac{V'(x_s^+) + 1}{\alpha_s^2} - 2\theta_s^c \frac{\alpha_s - \alpha_{s-1}}{(x_s-x_{s-1})\alpha_s^3}  =\frac{V'(x_s^+)}{\alpha_s^2} - 2\theta_s^c \frac{\alpha_{s+1} - \alpha_{s}}{(x_{s+1}-x_{s})\alpha_s^3} \,,
\end{equation*}
or equivalently
\begin{equation}
\alpha_s = \frac{2\theta_s^c \left[ \alpha_{s-1}(x_{s+1}-x_s) + \alpha_{s+1} (x_s - x_{s-1})\right]}{2\theta_s^c (x_{s+1}-x_{s-1}) -(x_{s+1}-x_s) (x_s - x_{s-1})}\,.\label{eqn:alpha_s_iter}
\end{equation}
This is not a linear problem, as $\theta_s^c$ depends on $\alpha_s$ through $\theta_{s+1}^c, \theta_{s+1}^s$ in a non-linear way (Equations \ref{eqn:price_cont} and \ref{eqn:der_cont}). Starting with the algorithm described in Section \ref{sec:pdde_solution} to compute $\theta^c, \theta^s$, adopting a linear interpolation as initial guess for $\alpha_s$, we may however apply the following iteration 
\begin{itemize}
	\item Update $\alpha_s$ through Equation \ref{eqn:alpha_s_iter}.
	\item Update $\theta_s^c, \theta_s^s$ from  $\theta_{s+1}^c, \theta_{s+1}^s$ using this new $\alpha_s$ via Equations \ref{eqn:price_cont} and \ref{eqn:der_cont}.
 	\item Update $\theta_{s-1}^c, \theta_{s-1}^s$ from $\theta_{s-2}^c, \theta_{s-2}^s$ using this new $\alpha_s$ via Equations \ref{eqn:price_cont} and \ref{eqn:der_cont}.
	\item Recalculate $\rho_R, \rho_L$ to ensure the jump condition at $x_s$, via Equations \ref{eqn:rhor} and \ref{eqn:rhol}. Scale $\theta^c, \theta^s$ by the new $\rho_R, \rho_L$.
\end{itemize}
One iteration is good enough for practical purposes, three iterations is nearly exact (Figure \ref{fig:density_blackflat_lvg_zoom}). 



\subsection{Interpolation of multiple expiries}
In order to illustrate the difference between the two calibration strategies outlined in Section \ref{sec:calibration_multiple}, we consider the market data of \cite{kahale2004arbitrage} for options on the SPX500 index as of October 1995. The quality of the market date is not great, but it is a good illustration of what may happen in not-so-liquid markets. 

\begin{figure}[!h]
	\subfigure[\label{fig:variance_kahale_lvg1}Independent calibration.]{
		\includegraphics[width=0.48\textwidth]{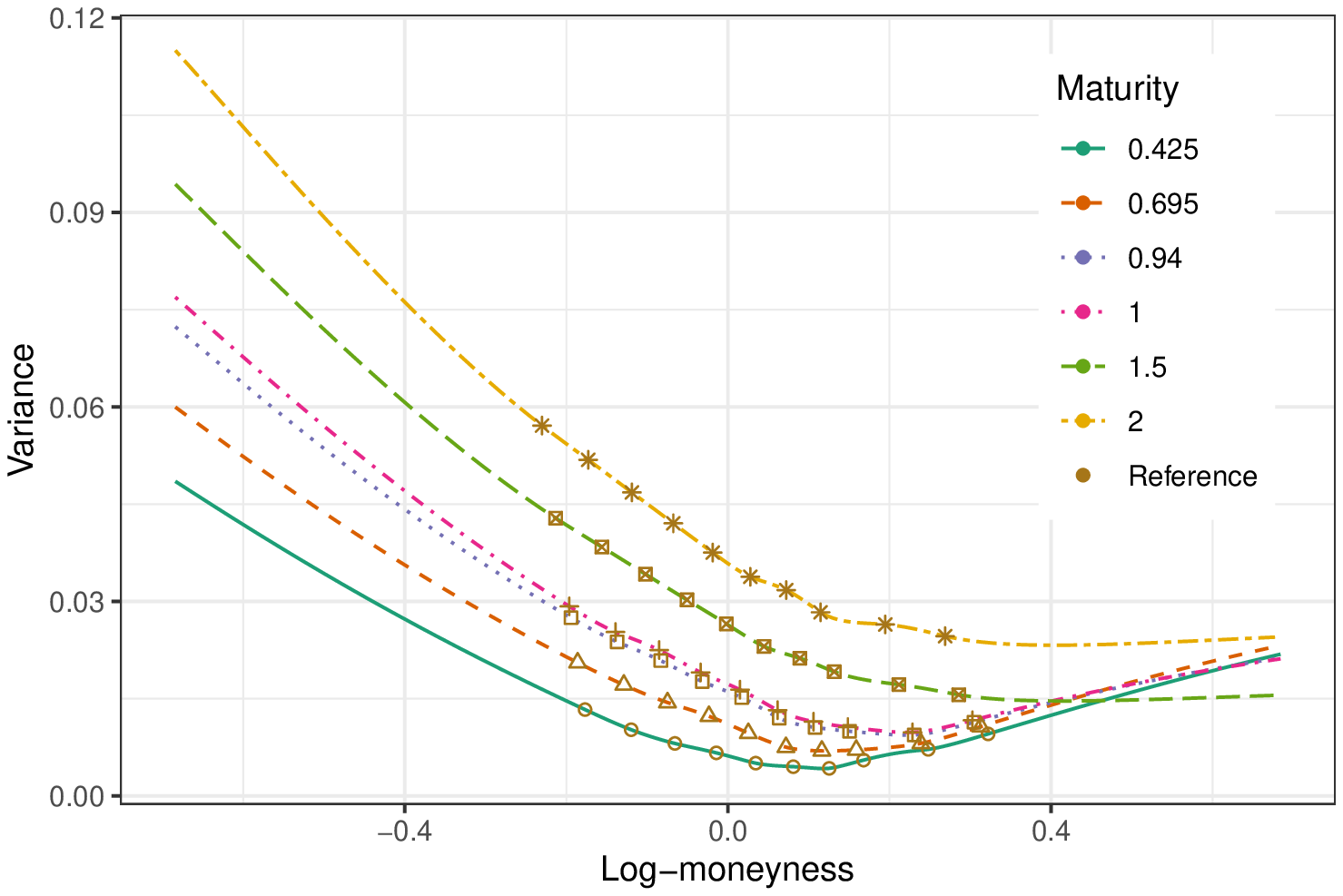}}
	\subfigure[\label{fig:variance_kahale_lvgm}Bootstrap calibration.]{
		\includegraphics[width=0.48\textwidth]{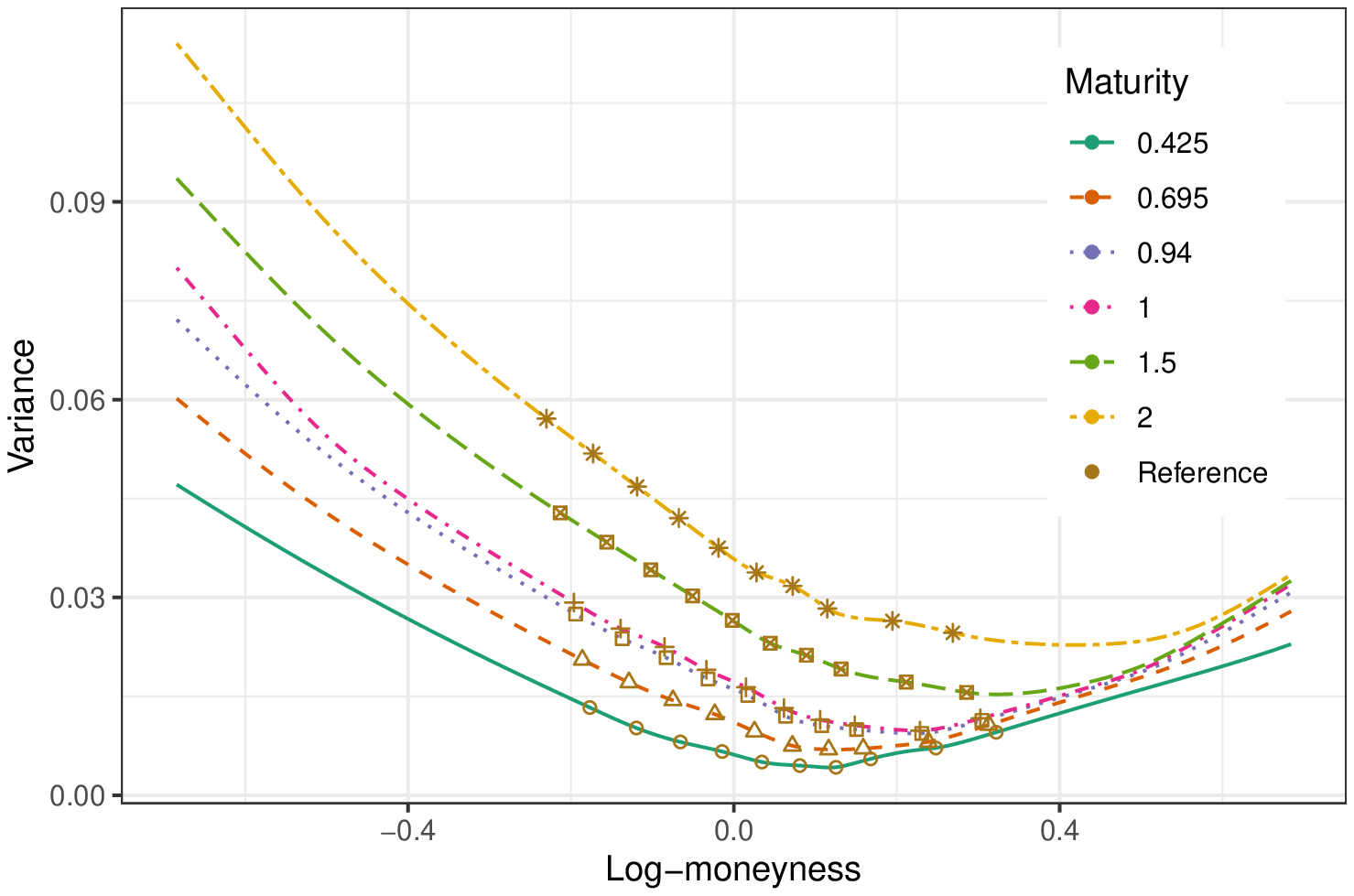} }
	\caption{Implied variance against log-moneyness with two different calibration strategies for the LLVG model, on SPX500 options as of October 1995.}
\end{figure}

If we calibrate the LLVG model to each option maturity $T_i$ independently, from $T=0$ to $T=T_i$, and plot the total variance $\sigma^2(T_i,x) T_i$ as a function of the log-moneyness $y = \ln \frac{x}{F(0,T_i)}$, where $\sigma$ is the calibrated LLVG model implied volatility and $F(0,T_i)=e^{(r-q)T_i}$ is the forward to maturity, with $r, q$ respectively the interest rate and dividend yield. It took around 0.4 ms to calibrate a single maturity and 2.2 ms to calibrate the full volatility surface. This is several orders of magnitude faster than the calibration time reported in \cite[Table 4]{carr2018expanded}. We notice that the lines for different maturities cross in the extrapolation part, around $y=0.4$ on Figure \ref{fig:variance_kahale_lvg1}.
 We thus have calendar spread arbitrages over there \citep{gatheral2006volatility}. If we calibrate the LLVG model in a bootstrap manner, from $T_{i-1}$ to $T_i$, using a linear interpolation of options prices at $T_{i-1}$, the lines do not cross anymore (Figure \ref{fig:variance_kahale_lvgm}). We however needed to be careful to add many\footnote{We chose 50 equidistant option prices, 20 may be sufficient.} option prices for the linear interpolation at $T_{i-1}$, otherwise, the resulting smile was not necessarily as smooth.
 
\begin{figure}[h]
 	\centering{
 		\includegraphics[width=.9\textwidth]{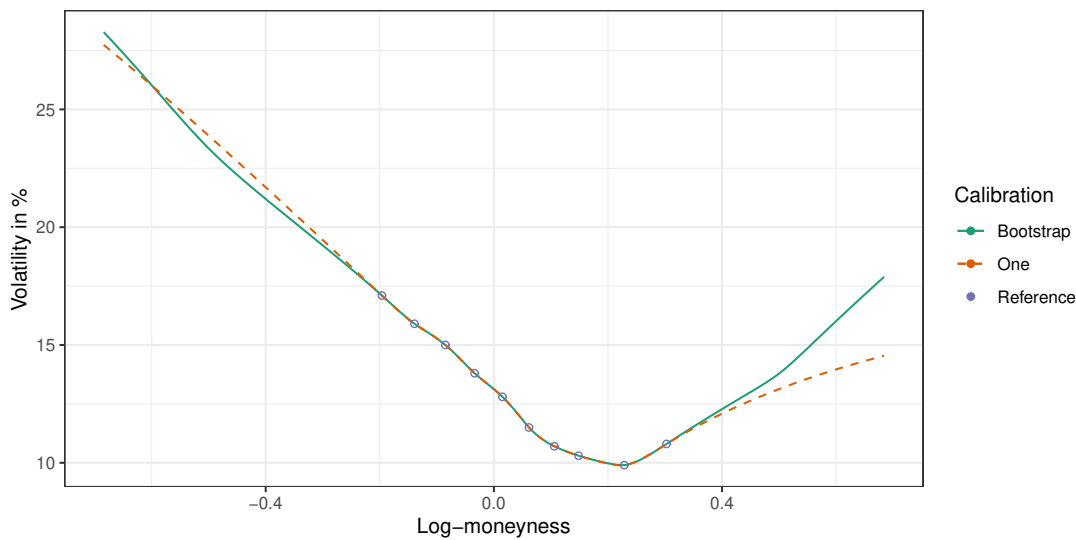}}
 	\caption{Implied volatility smile for the maturity $T_5=1$ year with the independent or bootstrap calibrations of the LLVG model.\label{fig:variance_kahale_lvg1m}}
 \end{figure}

 The implied volatility smiles are nearly the same within the interpolation range, even after bootstrapping many maturities (see Figure \ref{fig:variance_kahale_lvg1m}).

\section{Conclusion}
We have presented simple formulae, which allow to efficiently calibrate the local variance gamma model with a piecewise-linear representation of the local variance (LLVG model). In particular, we found the calibration to be orders of magnitude faster than what has been reported in \cite{carr2018expanded}. An exact interpolation of a small set of option quotes typically requires less than one millisecond.

For the calibration of a single maturity, the LLVG model possesses many advantages over the one-step finite difference technique of \citet{andreasen2011volatility}: it is faster, offers a continuous, smooth, interpolation, is more robust on the challenging examples of \citet{jackel2014clamping}, and there is no need to choose a proper discretization grid. We also have presented how to calibrate the LLVG model to multiple maturities, with a focus on avoiding calendar spread arbitrages. 

We proposed a specific regularization for the LLVG model, which leads to a smooth implied probability density, even when the options quotes used are noisy. We also put in evidence a flaw in the LLVG model if the variance function $a(x)$ is linear across the payoff discontinuity, or more generally of class $\mathcal{C}^1$. The one-step finite difference method suffers from the same issue. It may be solved by the proposed regularization, or via a simple iterative algorithm  without regularization.

We leave for further research an extension of the local variance gamma model with each piece of the type $a(x)= \frac{x^2}{q^2 x^2 + r^2}$. It is still analytically tractable  as the corresponding Dupire PDDE can be seen as a modified Bessel equation. 

Another area of research would be to improve the calibration technique for an exact interpolation. Although we have found augmented Gauss-Newton solvers like Levenberg-Marquardt to work well, can the calibration be reduced to a sequence of one-dimensional non-linear problems? Would it be more robust or more efficient?

\funding{This research received no external funding.}
\conflictsofinterest{The authors declare no conflict of interest.}
\externalbibliography{yes}
\bibliography{arbfree_interpolation}
\appendixtitles{no}
\appendix
\section{Relation between the weights in the least squares minimization of option prices and implied volatilities}\label{sec:weights}
We can find a weight $w_i$ that makes the solution similar to the one under the measures $E$ and $E_V$ by matching the gradients of each problem. We compare
\begin{equation*}
\sum_{i=1}^{m} 2 {w}_i^2  \frac{\partial C}{\partial \alpha}(\alpha, x_i)   \left( \hat{C}_i - C(\alpha, x_i) \right)\,,
\end{equation*} with
\begin{equation*}
\sum_{i=1}^{m} 2 {\mu_i}^2\frac{\partial \sigma}{\partial \alpha}(\alpha, x_i)  \left( \hat{\sigma_i} - \sigma(\alpha, x_i)\right)\,.
\end{equation*}
As we know that
$ \frac{\partial C}{\partial \alpha} =  \frac{\partial \sigma}{\partial \alpha} \frac{\partial C}{\partial \sigma} $,
we approximate $\frac{\partial C}{\partial \sigma}$ by the market Black-Scholes Vega, the term 
$\left( \hat{C}_i - C(\alpha, x_i) \right) $ by$\frac{\partial C}{\partial \alpha} (\alpha_{opt}-\alpha)$, and
$ \left( \hat{\sigma_i} - \sigma(\alpha, x_i) \right) $ by $\frac{\partial \sigma}{\partial \alpha} (\alpha_{opt}-\alpha) $
to obtain
\begin{equation}
{w}_i \approx \frac{1}{ \frac{\partial \hat{C}_i}{\partial \hat{\sigma_i}} } {\mu_i}\,.
\end{equation} 

In practice the inverse Vega needs to be capped to avoid taking into account too far out-of-the money prices, which won't be all that reliable numerically and we take
\begin{equation}
w_i = \min\left(\frac{1}{\nu_i}, \frac{10^6}{F} \right)\mu_i\,,
\end{equation}
where $\nu_i= \frac{\partial c_i}{\partial \sigma}$ is the Black-Scholes Vega corresponding the market option price $c_i$.

\section{De-arbitraging the market option prices: quadratic programming problem}\label{sec:arbitrage_free_qp}
Let $(c_i)_{i=1,...,m}$ be the undiscounted market call option prices on an asset $S$, with strike $y_i$ and forward price to maturity $T$ given by $f=\mathbb{E}[S(T)]$. The closest arbitrage-free option prices $\tilde{c}$ are the solution the following quadratic programming problem \cite{lefloch2019model}:
\begin{equation}
\tilde{c} = \argmin_{z \in \mathbb{R}^{n+1}} \left\lVert W \cdot (z- c)\right\lVert^2_2
\end{equation}
subject to 
\begin{equation}
-1< \frac{z_i-z_{i-1}}{y_i - y_{i-1}} < \frac{z_{i+1}-z_{i}}{y_{i+1} - y_{i}} <0 \,, \textmd{ and } z_i > \max(f-y_i,0)\,, \textmd{ for } i=2,...,m-1\,,
\end{equation}
where $W$ is a diagonal matrix of weights. For equal weights, $W$ is the identity matrix $I_{m}$. We can include information on the bid-ask spread, for example by taking $w_i$ to be the inverse of the bid-ask spread at strike $y_i$.

We have 
\begin{align*}
\left\lVert W \cdot (z- c)\right\lVert^2_2 &= z^T W^T W z - 2(W^T W c)^T z + (W c)^T W c\,. 
\end{align*}
The minimization problem can thus be formulated as a quadratic programming problem:
\begin{equation}\label{eqn:call_qp}
\tilde{c} =\argmin_{z \in \mathbb{R}^m, G z \leq h}  \frac{1}{2} z^T Q z + q^T z
\end{equation}
with  
\begin{align*}
Q &=W^T W\,, \\
q &= - W^T W c\,,\\
\end{align*} 
and the elements $G_{i,j}$ of the matrix $G$, that specifies the linear constraints in (\ref{eqn:call_qp}), are
\begin{equation*}
G_{i,i-1} = -\frac{1}{y_i-y_{i-1}}\,,\quad G_{i,i} = \frac{1}{y_i-y_{i-1}}+\frac{1}{y_{i+1}-y_i}\,,\quad
G_{i,i+1} = -\frac{1}{y_{i+1}-y_i} \,,
\end{equation*}
for $i=2,...,m-1$, and  
\begin{align*}
G_{1,1} = \frac{1}{y_2-y_1}\,,&\quad G_{1,2}=-\frac{1}{y_2-y_1}\,,\\
G_{m,m} = \frac{1}{y_m-y_{m-1}}\,,&\quad G_{m,m-1}=-\frac{1}{y_{m}-y_{m-1}}\,.\\
\end{align*}
and the vector $h$ by $h_1=1.0-\epsilon$, $h_i=-\epsilon$ for $1<i\leq m$.
The lower bound constraint translates to
\begin{equation*}
G_{m+i+1,i} = -1\,,\quad h_{m+i+1} = -\max(f-y_i,0)-\epsilon\,,\quad \textmd{ for } i=1,...,m\,.
\end{equation*}
The constant $\epsilon$ defines a maximum acceptable slope and ensures that the call prices are strictly convex.
\section{Market data}\label{sec:market_data_jackel}

\begin{table}[h]
	\centering{
		\caption{Black-Scholes implied volatilities against moneyness $\frac{x}{X(0)}$ for an option of maturity $T=5.0722$, examples 1 and 2 of \cite{jackel2014clamping}.}
		\begin{tabular}{rrr}\toprule
			\multicolumn{1}{c}{Moneyness} & \multicolumn{1}{c}{Volatility (Case I)} &  \multicolumn{1}{c}{Volatility (Case II)} \\ \midrule
			0.035123777453185 & 0.642412798191439 & 0.649712512502887 \\ 
			0.049095433048156 & 0.621682849924325 & 0.629372247414191\\ 
			0.068624781300891 & 0.590577891369241 & 0.598339248024188 \\ 
			0.095922580089594 & 0.553137221952525 & 0.560748840467284 \\ 
			0.134078990076508 & 0.511398042127817 & 0.518685454812697 \\ 
			0.18741338653678 & 0.466699250819768 & 0.473512707134552 \\ 
			0.261963320525776 & 0.420225808661573 & 0.426434688827871 \\ 
			0.366167980681693 & 0.373296313420122 & 0.378806875802102 \\ 
			0.511823524787378 & 0.327557513727855& 0.332366264644264 \\ 
			0.715418426368358 & 0.285106482185545 & 0.289407658380454\\ 
			1 & 0.249328882881654 & 0.253751752243855\\ 
			1.39778339939642 & 0.228967051575314 & 0.235378088110653 \\ 
			1.95379843162821 & 0.220857187809035 & 0.235343538571543 \\ 
			2.73098701349666 & 0.218762825294675 & 0.260395028879884 \\ 
			3.81732831143284 & 0.218742183617652 & 0.31735041252779 \\ 
			5.33579814376678 & 0.218432406892364 & 0.368205175099723 \\ 
			7.45829006788743 & 0.217198426268117 & 0.417582432865276 \\ 
			10.4250740447762 & 0.21573928902421 & 0.46323707706565 \\ 
			14.5719954372667 & 0.214619929462215 & 0.504386489988866 \\ 
			20.3684933182917 & 0.2141074555437 & 0.539752566560924 \\ 
			28.4707418310251 & 0.21457985392644  & 0.566370957381163\\ \bottomrule
		\end{tabular}
		\label{tbl:jackel_clamping_1}
	}
\end{table}

\begin{table}[h]
	\caption{Black-Scholes implied volatilities for SPX500 options in October 1995 from Nabil Kahale in \cite{kahale2004arbitrage}. The spot price was $S=590$ and the interest and dividend rates were $r=6\%$ and $q=2.62\%$ for each expiry $T$. \label{tbl:spx500_1995}}
	\centering{
		\begin{small}
			\begin{tabular}{@{}rrrrrrrrrrr@{}}
				\toprule
				$T$ & 85\% & 90\% & 95\% & 100\% & 105\% & 110\% & 115\% & 120\% & 130\% & 140\% \\ \midrule
				0.175 & 0.190 & 0.168 & 0.133 & 0.113 & 0.102 & 0.097 & 0.120 & 0.142 & 0.169 & 0.200 \\ 
				0.425 & 0.177 & 0.155 & 0.138 & 0.125 & 0.109 & 0.103 & 0.100 & 0.114 & 0.13 & 0.150 \\ 
				0.695 & 0.172 & 0.157 & 0.144 & 0.133 & 0.118 & 0.104 & 0.100 & 0.101 & 0.108 & 0.124 \\ 
				0.94 & 0.171 & 0.159 & 0.149 & 0.137 & 0.127 & 0.113 & 0.106 & 0.103 & 0.100 & 0.110 \\ 
				1 & 0.171 & 0.159 & 0.150 & 0.138 & 0.128 & 0.115 & 0.107 & 0.103 & 0.099 & 0.108 \\ 
				1.5 & 0.169 & 0.160 & 0.151 & 0.142 & 0.133 & 0.124 & 0.119 & 0.113 & 0.107 & 0.102 \\ 
				2 & 0.169 & 0.161 & 0.153 & 0.145 & 0.137 & 0.13 & 0.126 & 0.119 & 0.115 & 0.111 \\ 
				3 & 0.168 & 0.161 & 0.155 & 0.149 & 0.143 & 0.137 & 0.133 & 0.128 & 0.124 & 0.123 \\ 
				4 & 0.168 & 0.162 & 0.157 & 0.152 & 0.148 & 0.143 & 0.139 & 0.135 & 0.13 & 0.128 \\ 
				5 & 0.168 & 0.164 & 0.159 & 0.154 & 0.151 & 0.148 & 0.144 & 0.14 & 0.136 & 0.132 \\ \bottomrule
			\end{tabular}
		\end{small}
	}
\end{table}

\end{document}